# Multi-modal electron microscopy study on decoherence sources and their stability in Nb based superconducting qubit

Jin-Su Oh,[1] Xiaotian Fang,[1] Tae-Hoon Kim,[1] Matt Lynn,[1] Matt Kramer,[1] Mehdi Zarea,[2] James A. Sauls,[2] A. Romanenko,[3] S. Posen,[3] A. Grassellino,[3] Cameron J. Kopas,[4] Mark Field,[4] Jayss Marshall,[4] Hilal Cansizoglu,[4] Joshua Y. Mutus,[4] Matthew Reagor,[4] and Lin Zhou[1*]

[1]Ames Laboratory, Ames, Iowa 50011, United States

[2]Department of Physcis and Astronomy, Northwestern University, Evanston, IL 60208, United States

[3]Fermi National Accelerator Laboratory, Batavia, IL 60510, United States

[4]Rigetti Computing, Berkeley, California 94710, United States

*Corresponding author: linzhou@ameslab.gov

**Abstract**

Niobium is commonly used for superconducting quantum systems as readout resonators, capacitors, and interconnects. The coherence time of the superconducting qubits is mainly limited by microwave dissipation attributed to two-level system defects at interfaces, such as the Nb/Si and Nb/air interface. One way to improve the Nb/air interface quality is by thermal annealing, as shown by extensive studies in 3D superconducting radio frequency (SRF) cavities. However, it is unclear how the microstructure and chemistry of the interface structures change during heat treatment. To address this knowledge gap, we comprehensively characterized Nb films deposited on Si wafers by physical vapor deposition, including (1) an Nb film from a transmon and (2) an Nb film without any patterning step, using an aberration-corrected transmission electron microscope. Both Nb films exhibit columnar growth with strong [110] textures. There is a double layer between the Nb film and Si substrate, which are amorphous niobium silicides with different Nb and Si concentrations. After *in-situ* heating of the heterostructure at 360°C inside the microscope, the composition of the double layers at the Nb-Si interface remains almost the same despite different thickness changes. The initial amorphous niobium oxide layer on Nb surface decomposes into face-centered cubic Nb nanograins in the amorphous Nb-O matrix upon heating.

**Introduction**

Superconducting materials are promising candidates for solid-state quantum computers.[1–6] Nb thin film deposited on Si substrate is commonly used for readout resonators, capacitors, and interconnects in quantum circuits,[7] because Nb has advantages of a high superconducting gap and good surface quality for advanced lithographic patterning in the device.[8] Noise and decoherence in superconducting quantum processors mainly attest to the two-level system (TLS) at interfaces of the device, including metal-air (MA), metal-substrate (MS), substrate-air (SA) interfaces.[9,10] Thus, it is of great importance to understand the microscopic origin of the TLS presented at different interfaces. In particular, the $NbO_x$ ($0<x\leq2.5$) at the MA interface is considered as one of the lossiest parts of the superconducting quantum computing architectures.[9] The native niobium oxide is a detrimental cause for the dissipation of microwave, and the $NbO_x$ clusters and $Nb-NbO_x-Nb_2O_5$ interfaces degrade the superconductivity of niobium.[10–14]

Niobium pentoxide ($Nb_2O_5$) is a dominant form of the native oxide on the Nb surface. Previous studies show that the MA interface evolves as air-$Nb_2O_5$-$NbO_2$-$NbO$-$Nb$ bulk, or even $NbO_y$ ($y<1$) between NbO and bulk Nb under ambient conditions.[15–17] Particularly, $Nb_2O_5$ is considered as the main reason for quality factor degradation in superconducting Nb cavities. A thicker $Nb_2O_5$ surface film increases the microwave dissipation attributed to the TLS present in $Nb_2O_5$.[18] One way to suppress the TLS losses by $Nb_2O_5$ is thermal annealing in an ultra-high vacuum. Recent secondary ion mass spectrometry experiments on Nb cavities found that the $Nb_2O_5$ layer was completely disassociated after ~400°C heat treatment in a vacuum for 30 minutes, drastically improving the cavity quality factor.[19] These improvements are attributed to the decomposition of $Nb_2O_5$ into NbO, or even $Nb_2O$, which have metallic characteristics.[12,15,20–22] However, most studies on the decomposition of niobium oxide have been carried out using surface analysis techniques, which mainly focus on the chemistry of the niobium oxides.[12,15,20–22] The detailed microstructural evolution of the oxide layer and the MA interface under heat treatment is still unclear. In addition, the Nb-Si interface can be a source of microstructural imperfections, which may include non-superconducting metallic layers or additional TLS.

In this study, we investigated the Nb film fabricated by Rigetti Computing using aberration-corrected *in-situ* (scanning) transmission electron microscopy ((S)TEM) and spectroscopy.[23] A Nb-rich and a Si-rich amorphous Nb-Si layer were observed at the Nb-Si interface. In addition, we detail the structural differences of the Nb-Si MS and the Nb-O MA interfaces before and after *in-situ* heating experiment. We discovered that the initial $NbO_x$ layer decomposed into face-centered cubic (FCC) Nb nanograins embedded in the amorphous $NbO_x$ matrix. The thickness of the amorphous layers at Nb-Si interface was increased as well after heating. This investigation provides insight into the potential microscopic factors influencing coherence time at the MA and MS interfaces in Nb-based superconducting quantum systems.

**Experimental details**

The Nb films were deposited through physical vapor deposition by Rigetti Computing for producing their transmon qubits.[23] We investigated two types of samples: (1) Nb film from a transmon and (2) Nb film without any patterning step. Similar microstructure in the Nb film on Si was observed in both samples.

Transmission Kikuchi diffraction (TKD) was carried out in a scanning electron microscope (Teneo, Thermo Fisher Scientific, Ltd) equipped with an electron backscatter diffraction detector system at 20 kV. A probe current of 6.4 nA and a working distance of 8 mm was used during data acquisition. Cross-section and plan-view TEM samples were prepared by a focused ion beam with a gas injection system (Helios, Thermo Fisher Scientific Ltd.). The Nb thin film's surface was protected with a layer of photoresist before moving to the FIB chamber for cross-sectional TEM sample preparation. The TEM samples were thinned to electron beam transparency by a $Ga^+$ ion beam from 30 to 2 kV. The sample for in-situ heating was then dipped in acetone to remove the photoresist on top of the oxide. The TEM samples were investigated by an aberration-corrected TEM (Titan Cube, Thermo Fisher Scientific Ltd.) at 200 kV. A high angle annular dark-field (HAADF) detector was used for dark-field imaging in STEM mode with a convergent semi-angle and a collection semi-angle of 18 mrad and 74-200 mrad, respectively. *In-situ* heating experiment was carried out with an *in-situ* heating TEM holder (Model 628, Gatan, Inc). The TEM sample was directly heated to 280 °C followed by a gradual temperature increase to 360 °C within 30 min with a 20°C step size. The temperature was measured at the holder cup. Energy-dispersive X-ray spectroscopy (EDS) and electron energy-loss spectroscopy (EELS) studies were carried out with probe currents of 150 nA and 40 nA, respectively. Dual EELS was performed to acquire all-electron energy-loss (ELL) spectra. The plural scattering effect of all the raw ELL spectra was removed via Fourier-ratio deconvolution. Atomic models and corresponding simulated electron diffraction patterns were obtained using CrystalMaker software.

**Results and Discussion**

**Overall microstructures of the Nb film**

The Nb film has a polycrystalline structure with a preferred columnar growth along the [110] crystallographic direction. Figure 1 shows a cross-section TEM micrograph from the Nb resonator in the transmon. The Nb film has a thickness of ~170 nm with a high density of defects. The lower-right inset is a selected area electron diffraction pattern (SAEDP) obtained from the yellow circle indicated region (Figure 1a). Although few grains are selected in Figure 1a, the dominant (110) diffraction spots confirm the preferred growth of Nb film along the [110] direction on the Si(100) substrate. The columnar grains are better visualized in dark-field imaging conditions, as shown in Figure 1b, by using the dotted red circled spot in Figure 1a for imaging. The columnar grains have a lateral size of ~25 to ~55 nm. In addition, a plan-view TEM micrograph of the continuous Nb film was shown in Figure 1c. The upper-right inset is an inverse pole figure map from the plan-view sample. The Nb film's normal tends to align to the [110] direction, which is consistent with the cross-sectional TEM image in Fig. 1a and confirms the existence of fibrous texture in the Nb film.[24] In addition, SAEDP acquired from a large area of Nb plan-view sample shows broad arcs for most of the diffraction spots (Figure 1d), which are formed by the intersection of inclined reciprocal rings with the Ewald sphere.[25,26] Confirming the Nb film's [110] texture also has a high degree of mosaicity.

**Structure and thermal stability of the Nb/Si interface**

Close analysis at the Nb/Si interface shows no epitaxial relationship between Si and Nb. Instead, there are two layers between the Nb film and the Si substrate, as demonstrated by the high-resolution HAADF-STEM micrograph in Figure 2a. Figure 2b and c are EDS elemental maps of Si and Nb, respectively. It should be noted that the chemistry changes across the interface has a continuous profile, not a step function. The compositions of Nb and Si for each region are averaged compositions sectioned by white dashed line, as shown in Figure 2b and c. The first layer marked as layer 1 in Figure 2a is amorphous and ~ 2 nm thick. The composition of Nb and Si of layer 1 is close to 6:5. The second layer (layer 2) is ~ 4 nm thick and appears crystalline but with higher contrast than the Si substrate in the HAADF-STEM image. This is due to a higher amount of Nb in the second layer. Based on the irregular surface morphology of the Si substrate, as outlined by the yellow dashed line in the HAADF-STEM micrograph at atomic resolution in Figure 2d, the contrast of Layer 2 is due to the overlapping of an amorphous interface layer and the Si substrate along the electron beam direction. Energy-loss near-edge structures (ELNES) of the Nb-M edges and Si-L edge at the interface from the points denoted in Figure 2d are shown in Figure 2e and f. The Nb $M_{23}$ edge clearly shows a shoulder peak close to the Nb substrate (Figure 2e), which is evidence of the Nb forming partial valence states.[27,28] Likewise, the Si $L_{23}$ edge changes across the interface. A peak is observed at 102 eV from spot 10, consistent with crystalline Si observed in the HADDF image.[29] Closer to the Nb region, there is a noticeable peak shifting to higher energies (see the small black arrow). However, at the same time, the intensity of those peaks significantly decreases. This decrease is attributed to increased amorphous Si volume fraction in the interface.[29] From spots 6 to 3, no apparent peak is detected at ~102 eV, except the edge-onsets of Si $L_{23}$, which indicates only amorphous Si, in good agreement with the HAADF-STEM micrograph (see Figure 2d). Therefore, we conclude that layers 1 and 2 are amorphous niobium silicide, but layer 1 has a distinctly higher Nb content than layer 2.

After the TEM sample was heated to 360 °C for ~30 min, the thickness of both interface layers increased. The first layer (layer 1) grew from ~2 nm to ~4 nm, and the second layer from ~4 nm to ~5.5 nm, as shown in Figure 3a. Dominant growth of layer 1 compared to layer 2 perhaps originates from the different interdiffusion coefficients of Nb and Si. Prasad and Paul argued that the diffusion rate of Si is faster than that of Nb in Nb-Si system.[30] It is worth noting that the composition of the two interface layers remains the same after heating despite the change of layer thickness (see Figure 3b and c). Likewise, a spatially resolved EELS study shows similar chemistry of Nb and Si across the interface as before heating. The Nb $M_{23}$ edge clearly shows a shoulder feature when the probed spot gets into layer 1, as marked by a small black arrow in Figure 3e. A positive shift of Si $L_{23}$ edge is observed in layer 1, as marked by a small black arrow in Figure 3f, indicating an onset formation of the silicide. The similarity in the composition of the two layers and consistency in the EELS suggests that the amorphous structures with measured chemistry are thermodynamically favored under our experimental conditions. Amorphous Nb–Si alloys can be formed using a number of rapid quenching techniques,[31–33] including sputtering.[34] Rapid quenching of Nb-near the 18.7 at.% Si eutectic can form a metallic glass.[35] It is well established that metallic glasses form in the vicinity of deep eutectics and favor composition near the deepest decent,[36] and the compositions here fulfill this criteria, being close to the eutectics at 57 and 98% Si.[37]

**Structure and thermal stability of the niobium oxide layer**

The native oxide layer on Nb metal leads to a dielectric loss in the superconducting qubit circuits [9] and 3D SRF cavities.[18,19] This MA interface has the highest loss tangent among the MS, MA, and SA interfaces in 2D devices.[9] The loss tangent of the native niobium oxide in the relevant quantum regime due to TLS has been directly measured inside 3D SRF cavities to be as high as δ~0.1.[19] Furthermore, thermal annealing at ~300-400 °C has been demonstrated in 3D SRF cavities to dramatically improve the quality factors in quantum regime[19] by dissolving the native niobium oxide layer hosting the majority of TLS. X-ray photoelectron spectroscopy has been conducted to understand the annealing effect on the MA interface chemistry.[12,15,20–22] However, the microstructural evolution of the NbO$_x$ is still not fully understood. We studied the thermal stability of the Nb oxide layer using an *in-situ* heating TEM holder, and the base vacuum of the microscope column is 5×10$^{-5}$ mTorr. Microstructures of the same regions (marked by white lines) around the NbO$_x$-Nb interface before and after heating are shown in Figure 4a and b, respectively. The original NbO$_x$ layer is amorphous with a wavy surface and a thickness of ~5-13.4 nm. After the TEM sample is heated to 360 ˚C for ~30 min, the thickness of the initial amorphous NbO$_x$ layer is reduced to ~3-7.2 nm. The film is also partly transformed to a crystalline state, as shown by yellow arrows in Figure 4c and d. Moreover, the oxygen to niobium ratio decreased from 1.8 to 0.6 inside the NbO$_x$ layer after heating. Figure 4h and i indicate that the NbO$_x$ layer is significantly depleted in oxygen. In addition, the oxygen content in the Nb bulk region declines after heating (see Figure 4e-i), which may be due to the decomposition of NbO$_x$ on the TEM lamella surfaces. It should be noted that there is no noticeable morphology change at the original NbO$_x$-Nb interface during heating, indicating that the primary mechanism of NbO$_x$ thickness reduction is decomposition of the NbO$_x$ under our experimental condition.

Surprisingly, the newly formed crystalline phase after heating can only be indexed as FCC Nb. Figure 5a and b are high magnification TEM and atomic resolution HAADF-STEM micrographs of Nb-NbO$_x$ interface after heat treatment, respectively. The original amorphous NbO$_x$ layer decomposes to nanocrystalline grains intermixed with some residual amorphous NbO$_x$. While BCC structure is the thermodynamically stable phase at room temperature in bulk or sputtered Nb, FCC Nb may form with fairly high oxygen content. When BCC niobium is viewed along the [111] zone axis, d-spacing of (110) is 0.233 nm, and all angles between adjacent {110} diffraction spots should be 60˚. Fast-Fourier transformation is carried out for the grain denoted by red dashed box region in Figure 5b. The d-spacing of two plane groups are measured as 0.259 nm and 0.224 nm, which are close to $d_{(111)}$ = 0.244 nm and $d_{(200)}$ = 0.212 nm of FCC niobium for the lattice constant is a = 4.22 Å, respectively. Moreover, the angles measured from adjacent planes show 70.7 and 54.4 degrees, deviating largely from the BCC structure. For FCC niobium projected along the [011] direction, angles between (-111) and (-1-11) planes and between (-1-11) and (020) planes should be 70.53° and 54.74°, respectively, close to the measured angles, with lattice distortion and errors taken into account. The difference for niobium in atomic models between BCC and FCC from the projections mentioned above is clearly shown in Figure 5c and d. Moreover, corresponding simulated electron diffraction patterns are presented in Figure 5e and f. It shows that the diffraction pattern of the FCC niobium looks like it is stretched slightly up and down compared to BCC niobium. The formation of nanocrystalline FCC Nb can be explained in terms of free volume. When the grain size of Nb is below 10 nm, the increasing excess free volume of Nb atoms induces a characteristic lattice expansion within grains, consequently leading to transformation to FCC.[38,39] Although FCC Nb has higher resistivity and lower critical temperature, $T_c$, than BCC Nb,[40] it has a metallic nature that could reduce dielectric loss at the interface. While the improvement of low field Nb SRF cavity performance when subjected to thermal annealing at 300-400 °C is shown to be driven by the dissolution of the $Nb_2O_5$ layer,[19] the effect of the formation of FCC niobium on microwave dissipation warrants further investigations. Moreover, FCC Nb may undergo polymorphic transformation to BCC structure by grain growth under additional annealing.

However, the question remains why a significant amount of oxygen is still detected in the previous amorphous $NbO_x$ layer after heating, even though the decomposed crystalline region is determined to be FCC Nb. To clarify the chemistry and electronic structure change of Nb and O at the interface region, we perform spatially resolved EELS. Figure 6a and d are representative bright-field TEM images before and after heating. White dotted lines in the TEM micrographs denote the same position between crystalline Nb bulk and amorphous $NbO_x$. Figure 6 b, c, e, and f show spatially resolved EEL spectra of Nb $M_{23}$ and O K edges collected from the colored spots in Figure 6a and d. There are significant positive shifts of Nb $M_3$ peaks from 363 eV to 365 eV from the $NbO_x$ surface towards the $NbO_x$/Nb interface (from line 1 to line 3), indicating that the valence state is +5, i.e., $Nb_2O_5$, which is the dominant oxide on the Nb surface.[12] For line 4, a slightly lower positive shift of Nb $M_3$ is observed at 364.25 eV, indicating there's $NbO_x$ (1≤x<2.5) like $NbO_2$ or NbO between $Nb_2O_5$ and bulk Nb. Likewise, ELNES of O K edge also denotes that $Nb_2O_5$ and $NbO_2$ are dominant in the region of line 2-5 in Figure 6c, which is based on a peak produced at 545 eV only from $Nb_2O_5$ and $NbO_2$, not from NbO.[27] Line 6-8 corresponding to bulk Nb, no shift in Nb $M_{23}$ peaks is observed. There is no significant amount of oxygen. We suspect that the small O K edges are mainly from the surface oxide of the TEM lamella, which is hard to avoid when the relative thickness (t/λ, where t is the thickness of the sample and λ is the mean free path of the fast electrons) of the TEM sample is under 0.5.[27] After heat treatment, there are still positive shifts of the Nb $M_3$ edge from 363 eV to 363.5 eV in the previous $Nb_2O_5$ region, as shown by lines of 1-3 in Figure 6e, confirming the existence of NbO. Figure 6f shows peak intensity of the O K edge is significantly reduced upon heating, which is in good agreement with EDS quantification of oxygen in Figure 4. In addition, those O K ELNES changed considerably from that of $Nb_2O_5$, as peak intensity declined at 545 eV. This change indicates that the decomposition of $Nb_2O_5$ into NbO and/or FCC Nb occurred. No significant differences in valence states are observed in the bulk Nb region. Based on the above result, we conclude that the decomposed region consists of FCC Nb and NbO.

The observed larger lattice parameter (a ≈ 4.44 Å) in the decomposed FCC niobium may be related to the existence of oxygen at its octahedral and tetrahedral interstitial sites. Assuming that the Nb has FCC Bravais lattice, we use DFT to determine the optimal value of the lattice constant corresponding to the minimum of total ground state energy.[41,42] For pure FCC Nb, the lattice constant is a = 4.22 Å. For fcc Nb there are 4 atoms in fcc lattice corresponding to a ground state atomic density of $\rho = 1/18.76$ Å$^{-3}$. Note that for a BCC lattice with 2 atoms per unit cell, the same optimization method leads to a ≈ 3.33 Å, which is equivalent of $\rho \approx 1/18.14$ Å$^{-3}$. The presence of oxygen impurity increases the average Nb lattice spacing. To this end we use a 2×2×2 unit cell with 8 Nb atoms and 1 oxygen impurity initially located at (0.2, 0.2, 0.2) from a Nb atom. The ground state energy for this supercell leads to a lattice constant of a = 4.32 Å. Figure 7 shows the increasing of Nb lattice constant in the presence of oxygen atoms. The optimal value of lattice constant seems to be more than 4.4 Å. We note that this is when all relative position of atoms are fixed. Thus, a zero force condition on all the atoms is not gauranteed. We use the Broyden-Flecher-Goldfarb-Shanno nonlinear optimization algorithm[43] employed by Quantum Espresso to find zero-force configuration of the lattice, which yields a = 4.32 Å.[41,42]

A schematic representing the oxide decomposition mechanism at ~360 °C is presented in Figure 8. The initial air-Nb interface evolved as air-$Nb_2O_5$-$NbO_x$ (1≤x<2.5)-Nb. The interface between Nb and amorphous $NbO_x$ (1≤x<2.5), indicated by the white dashed line, will likely remain unchanged during short time heating. The thickness of the original oxide layer was reduced through the decomposition of $Nb_2O_5$ and $NbO_x$ (1≤x<2.5) into NbO. At the same time, annealing induced the formation of FCC Nb with a size under ~10 nm.

**Conclusion**

The detailed microstructures and their evolution at elevated temperature in Nb films for superconducting qubits are carried out using *in-situ* aberration-corrected TEM imaging and spectroscopy (EDS and EELS). We discover that Nb film has a columnar structure with a lateral grain size of ~ 25-50 nm. The grains are highly textured along the [110] direction. Atomic resolution investigation of the Nb-Si interface shows double layers between the Nb film and Si substrate, which are amorphous niobium silicides with different Nb and Si concentrations. The amorphous surface $NbO_x$ layer is dominated by $Nb_2O_5$ with suboxide(s) close to the Nb metal region. After heat treatment at 360 °C for 30 min, the two layers at the Nb-Si interface show no chemistry change despite the thickness increase of both layers. The amorphous niobium oxide layer decomposed into nanocrystalline FCC Nb in the amorphous NbO matrix. Our comprehensive structural study of Nb thin films on Si substrates provides guidance for future superconducting qubit device optimization through interfacial scattering center and TLS minimization.

**Acknowledgments**

This work was supported by the U.S. Department of Energy, Office of Science, National Quantum Information Science Research Centers, Superconducting Quantum Materials and Systems Center (SQMS) under contract No. DE-AC02-07CH11359. All electron microscopy and related work were performed using instruments in the Sensitive Instrument Facility in Ames Lab. The Ames Laboratory is operated for the U.S. Department of Energy by Iowa State University under Contract No. DE-AC02-07CH11358

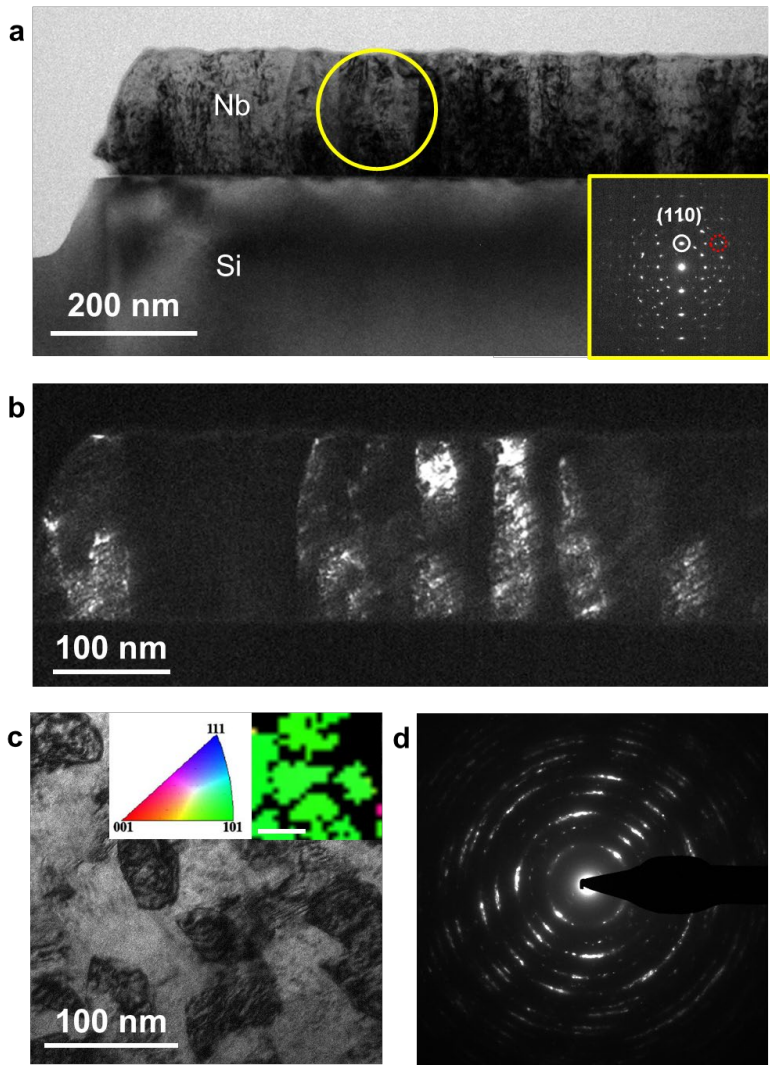

**Figure 1. Overall microstructure of Nb film. (a)** Low magnification bright-filed cross-sectional TEM image. Inset: Selected area electron diffraction pattern (SAEDP) from the region indicated by yellow circle in a. The Diffraction spot of (110) is parallel to the normal direction of Si substrate. **(b)** Dark-field TEM image shows columnar Nb grains using the diffraction spot indicated by the red dotted circle in a. **(c)** low magnification of plan-view TEM image of Nb film that shows grain size Nb. Inset: Inverse pole figure map of the normal direction of the Nb plan-view acquired by transmission Kikuchi diffraction, showing a strong [110] texture. Scale bar in the inset is 50 nm. **(d)** SAEDP from a large area of the Nb plan-view TEM sample. It shows arcs in the electron diffraction patterns indicative of a textured film.

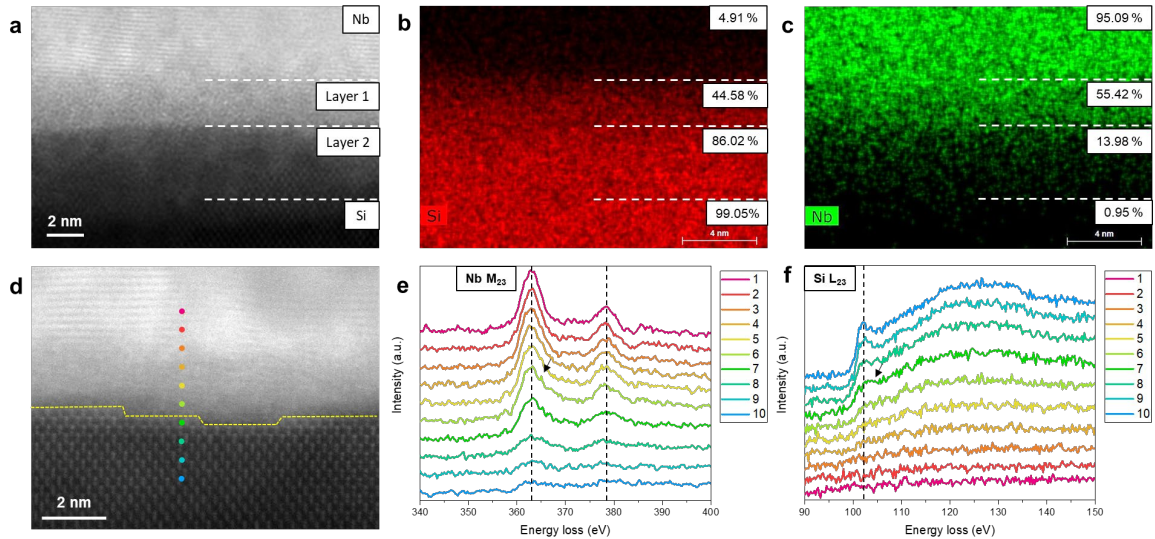

**Figure 2. Nb-Si interface. (a)** Atomic resolution HAADF-STEM micrograph showing the Nb-Si interface. Two layers are observed between the Nb and Si. Note that Layer 1 is amorphous. **(b, c)** EDS elemental maps of Si and Nb, respectively. Atomic percentages of the elements are shown at the upper right side of each region divided by white dashed lines. **(d)** Atomic resolution HAADF-STEM micrograph shows that the crystalline Si substrate surface is rough. The colored spots denote the positions in which EEL spectra are acquired. **(e, f)** Spatially resolved EEL spectra indicating Nb $M_{23}$ and Si $L_{23}$ ELNESs, respectively.

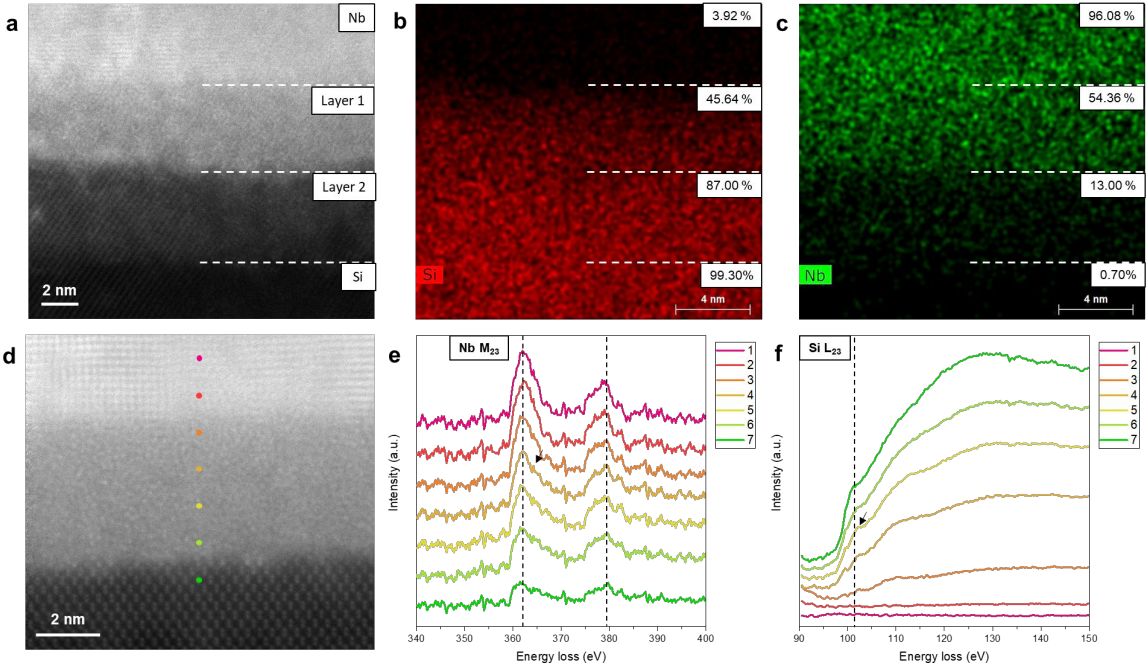

**Figure 3. Nb-Si interface after heating. (a)** Atomic resolution HAADF-STEM micrograph showing the Nb-Si interface. Two interface layers become thicker upon annealing. **(b, c)** Corresponding EDS elemental maps of Si and Nb, respectively. Atomic percentages of the elements are shown at the upper right side of each region divided by white dashed lines. **(d)** Atomic resolution HAADF-STEM micrograph. The colored spots denote the positions in which EEL spectra are acquired. **(e, f)** Spatially resolved EEL spectra indicating Nb $M_{23}$ and Si $L_{23}$ ELNESs, respectively.

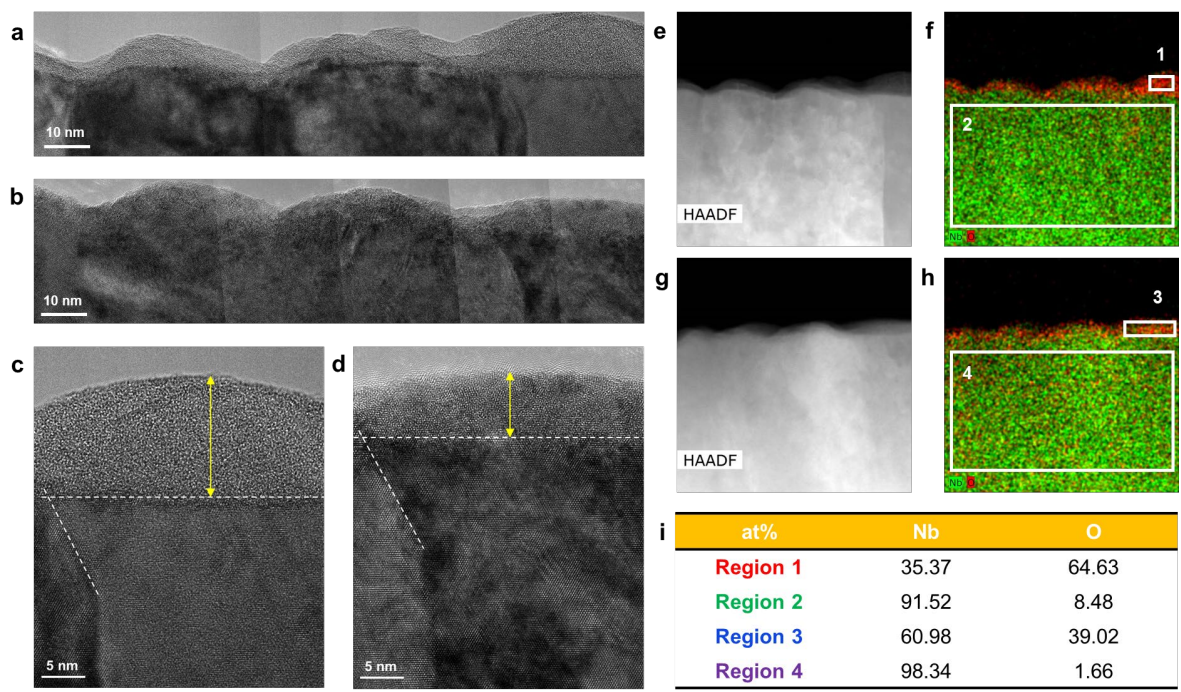

**Figure 4. Microstructure of Nb/NbOx interface before and after in-situ heating. (a, b)** cross-section TEM images of Nb/NbO$_x$ interface before and after in-situ heating, respectively. **(c, d)** High-resolution TEM images of Nb/NbO$_x$ interface before and after in-situ heating, respectively. White lines show the same location of the sample before and after heating. Note that the thickness of NbO$_x$ layer decreases after heating and nanocrystalline grains forms in the NbO$_x$ layer, which is indicated by yellow arrows. Location of the original interface almost unchanged. **(e, f)** HAADF-STEM image and corresponding EDS elemental maps before heating, respectively. **(g, h)** HAADF-STEM images and corresponding EDS elemental map after heating, respectively. **(i)** Atomic percentages of Nb and O in from the white boxed region with numbers in f and h. The oxygen concentration decreases after heating in the NbO$_x$ and Nb.

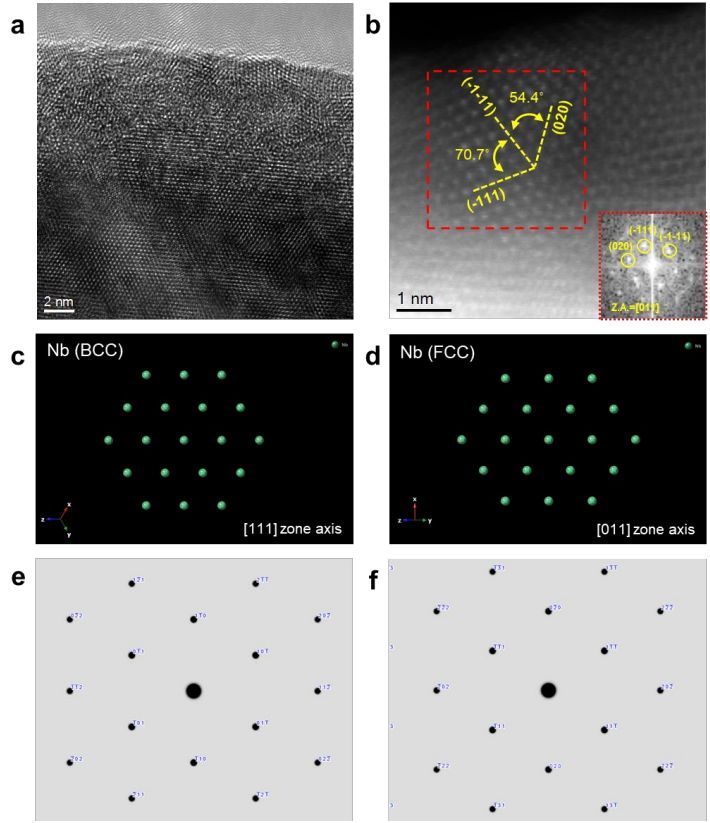

**Figure 5. Structure of the nanocrystalline Nb grain. (a)** High-resolution TEM micrograph of Nb/NbO$_x$ interface region after heating. **(b)** Atomic resolution HAADF-STEM micrograph of decomposed region. Red dashed box indicates a Nb grain with FCC structure. Inset: corresponding FFT of red dashed boxed region. **(c, d)** Atomic structures of BCC Nb along [111] direction and FCC Nb along [011] direction, respectively. **(e, f)** Simulated diffraction patterns of BCC Nb from [111] zone axis and FCC Nb from [011] zone axis, respectively.

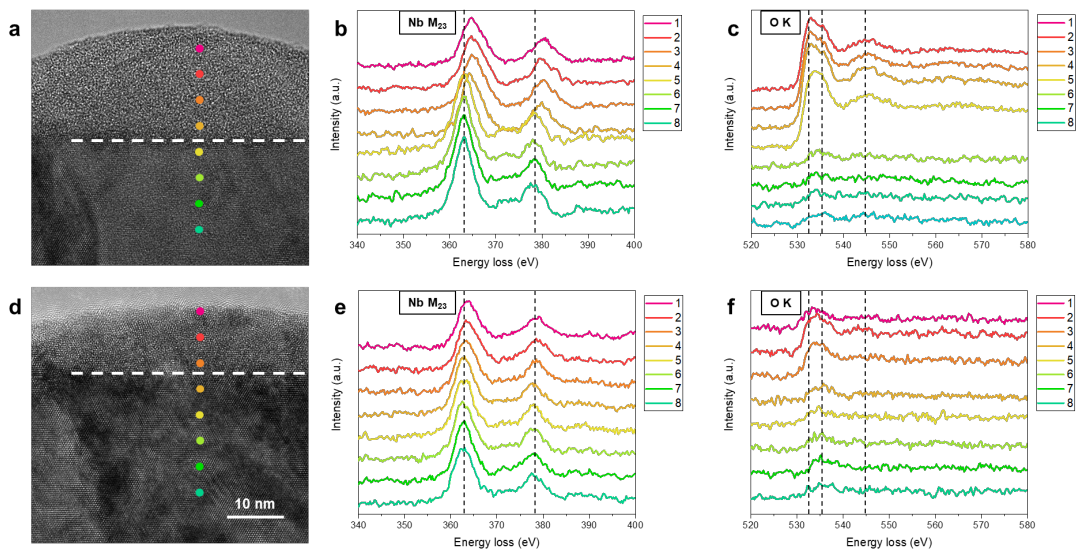

**Figure 6. Depth profile of EELS line scan of Nb/NbOx region before and after heating. (a)** bright-field TEM image of Nb/NbO$_x$ region before heating. The colored spots indicate positions of acquired EEL spectra. **(b, c)** Spatially resolved EEL spectra indicating Nb M$_{23}$ and O K ELNESs, respectively. **(d)** bright-field TEM image of NbO$_x$/Nb region after heating. The colored spots indicate positions of EEL spectra acquired. The white dotted line indicates the initial interface between Nb and NbO$_x$ is almost unchanged during heat treatment. **(e, f)** Spatially resolved EEL spectra indicating Nb M$_{23}$ and O K ELNESs, respectively. Note that there are still positive shifts of M$_{23}$ peaks in the region of previous NbO$_x$ layer although nanograins are observed in this region.

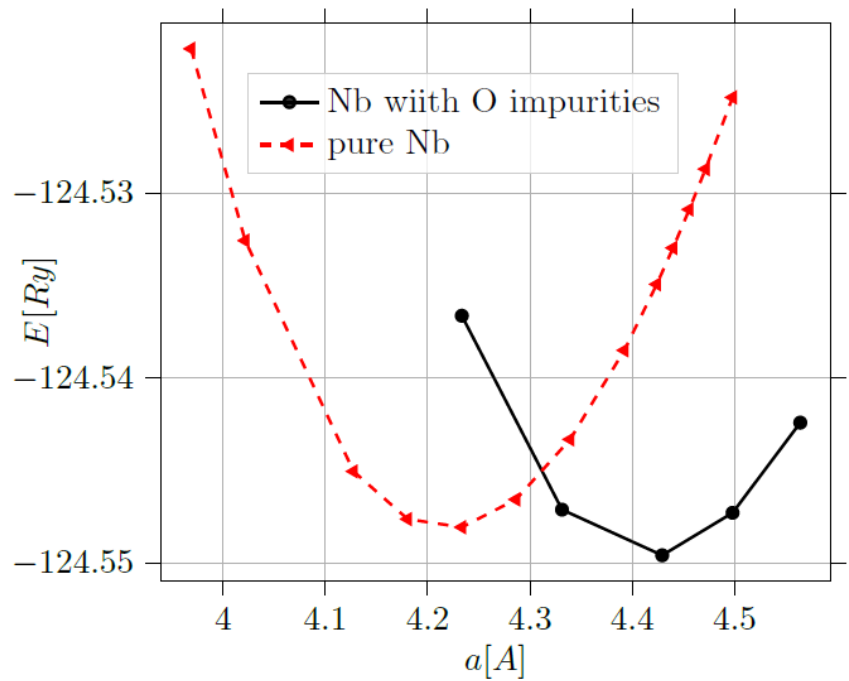

**Figure 7. The ground state energy versus the lattice constant.** For Nb with a single O impurity in a supercell with 8 Nb atoms. To compare with pure Nb the total ground state energy is divided by 8 then is shifted upward by 5:18 Ry.

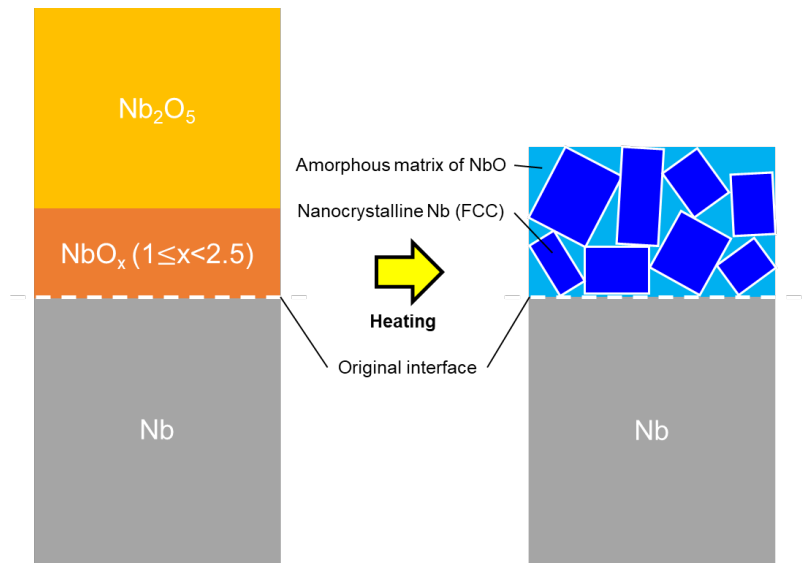

**Figure 8. Schematic showing decomposition of niobium oxide layer during heat treatment.** The initial amorphous layer which consists of $Nb_2O_5$, $NbO_x$ ($1 \leq x < 2.5$) decomposes into NbO and/or nanocrystalline FCC Nb.

the Local Atomic and Electronic Structure of Amorphous Oxidized Superconducting Niobium Films. 1–6.

(9) Woods, W.; Calusine, G.; Melville, A.; Sevi, A.; Golden, E.; Kim, D. K.; Rosenberg, D.; Yoder, J. L.; Oliver, W. D. Determining Interface Dielectric Losses in Superconducting Coplanar-Waveguide Resonators. *Phys. Rev. Appl.* **2019**, *12* (1), 1. https://doi.org/10.1103/PhysRevApplied.12.014012.

(10) Verjauw, J.; Potočnik, A.; Mongillo, M.; Acharya, R.; Mohiyaddin, F.; Simion, G.; Pacco, A.; Ivanov, T.; Wan, D.; Vanleenhove, A.; Souriau, L.; Jussot, J.; Thiam, A.; Swerts, J.; Piao, X.; Couet, S.; Heyns, M.; Govoreanu, B.; Radu, I. Investigation of Microwave Loss Induced by Oxide Regrowth in High-Q Nb Resonators. **2020**, *014018*, 1–18. https://doi.org/10.1103/PhysRevApplied.16.014018.

(11) Halbritter, J. Comment on "Direct Observation of the Superconducting Energy Gap Developing in the Conductivity Spectra of Niobium." *Phys. Rev. B - Condens. Matter Mater. Phys.* **1999**, *60* (17), 12505–12506. https://doi.org/10.1103/PhysRevB.60.12505.

(12) Ma, Q.; Rosenberg, R. A. Surface Study of Niobium Samples Used in Superconducting Rf Cavity Production. *Proc. IEEE Part. Accel. Conf.* **2001**, *2*, 1050–1052. https://doi.org/10.1109/pac.2001.986573.

(13) Halbritter, J. On the Oxidation and on the Superconductivity of Niobium. *Appl. Phys. A Solids Surfaces* **1987**, *43* (1), 1–28. https://doi.org/10.1007/BF00615201.

(14) Padamsee, H. *The Science and Technology of Superconducting Cavities for Accelerators*; 2001.

(15) Delheusy, M.; Stierle, A.; Kasper, N.; Kurta, R. P.; Vlad, A.; Dosch, H.; Antoine, C.; Resta, A.; Lundgren, E.; Andersen, J. X-Ray Investigation of Subsurface Interstitial Oxygen at Nb/Oxide Interfaces. *Appl. Phys. Lett.* **2008**, *92* (10), 8–11. https://doi.org/10.1063/1.2889474.23